# The mathematics of functional differentiation under conservation constraint

Tamás Gál

Department of Theoretical Physics, University of Debrecen, H-4010 Debrecen, Hungary

**Abstract:** The mathematics of *K*-conserving functional differentiation, with K being the integral of some invertible function of the functional variable, is clarified. The most general form for constrained functional derivatives is derived from the requirement that two functionals that are equal over a restricted domain have equal derivatives over that domain. It is shown that the *K*-conserving derivative formula is the one that yields no effect of *K*-conservation on the differentiation of *K*-independent functionals, which gives the basis for its generalization for multiple constraints. Connections with the derivative with respect to the shape of the functional variable and with the shape-conserving derivative, together with their use in the density-functional theory of many-electron systems, are discussed. Yielding an intuitive interpretation of *K*-conserving functional derivatives, it is also shown that *K*-conserving derivatives emerge as directional derivatives along *K*-conserving paths, which is achieved via a generalization of the Gâteaux derivative for that kind of paths. These results constitute the background for the practical application of *K*-conserving differentiation.

## I. Introduction

Constraints on distributions $\rho(x)$ in some space limit the changes of physical quantities depending on the distributions in many fields of physics. If functional differentiation with respect to $\rho(x)$ is involved in the given physical theory, a proper treatment of the constraints on $\rho(x)$ by the differentiation becomes necessary. In [1,2], for that proper treatment, the mathematical formula

$$\frac{\delta A[\rho]}{\delta_K \rho(x)} = \frac{\delta A[\rho]}{\delta \rho(x)} - \frac{f^{(1)}(\rho(x))}{K} \int \frac{f(\rho(x'))}{f^{(1)}(\rho(x'))} \frac{\delta A[\rho]}{\delta \rho(x')} dx' \qquad (1)$$

(with $\dfrac{\delta A[\rho]}{\delta \rho(x)}$ the unconstrained functional derivative of the functional $A[\rho]$) for constrained functional differentiation, under constraints of the form

$$\int f(\rho(x)) dx = K \qquad (2)$$

(where an explicit $x$-dependence of $f$ is allowed as well, though not denoted for simplicity), has been derived via the decomposition

$$\rho(x) = f^{-1}\left(\frac{K}{\int f(g(x'))dx'} f(g(x))\right) \qquad (3)$$

of the functional variable, applying the idea

$$\left(\frac{\delta A[\rho[g,K]]}{\delta g(x)}\right)_K \bigg|_{g=\rho} \equiv \frac{\delta A[\rho]}{\delta_K \rho(x)} . \qquad (4)$$

In this paper, the mathematical basics of $K$-conserving (or $K$-constrained) differentiation and of the formula Eq.(1) will be clarified, which is essential for practical applications and for generalizations for wider classes of constraints.

## II. Restricted functional derivatives

Two basic derivatives of a functional (or operator) are defined in functional analysis. The Fréchet derivative of a functional $A[\rho]$ at $\rho(x)$ is defined as a linear operator $F[\rho;.]$ that gives

$$F[\rho;\Delta\rho] = A[\rho+\Delta\rho] - A[\rho] + o[\rho;\Delta\rho] \qquad (5)$$

for any $\Delta\rho(x)$, with

$$\lim_{\Delta\rho\to 0} \frac{o[\rho;\Delta\rho]}{\|\Delta\rho\|} = 0 \;, \qquad (5a)$$

while an operator $G[\rho;.]$ is the Gâteaux derivative of $A[\rho]$ at $\rho(x)$ if it gives the so-called Gâteaux differential,

$$G[\rho;\Delta\rho] = \lim_{\varepsilon\to 0} \frac{A[\rho+\varepsilon\Delta\rho] - A[\rho]}{\varepsilon} \;, \qquad (6)$$

for any $\Delta\rho(x)$, and is linear and continuous [3]. (That $F[\rho;.]$, or $G[\rho;.]$, is the derivative of $A[\rho]$ is not denoted for simplicity). Both derivatives are defined uniquely, and their relation is characterized by the theorem that if the Fréchet derivative exists at a $\rho(x)$ then the Gâteaux derivative exists there as well and the two derivatives are equal, as can be seen by

$$G[\rho;\Delta\rho] = \lim_{\varepsilon\to 0} \frac{A[\rho+\varepsilon\Delta\rho] - A[\rho]}{\varepsilon} = \lim_{\varepsilon\to 0} \frac{F[\rho;\varepsilon\Delta\rho] - o[\rho;\varepsilon\Delta\rho]}{\varepsilon} =$$

$$\lim_{\varepsilon\to 0} \left\{ F[\rho;\Delta\rho] - \|\Delta\rho\| \frac{o[\rho;\varepsilon\Delta\rho]}{\varepsilon\|\Delta\rho\|} \right\} = F[\rho;\Delta\rho] \;. \qquad (7)$$

If constraint limits the changes $\Delta\rho(x)$ of $\rho(x)$, the existence of a general derivative is not needed, and the concept of a restricted derivative naturally arises through

$$F|_K[\rho;\Delta_K\rho] = A[\rho + \Delta_K\rho] - A[\rho] + o_K[\rho;\Delta_K\rho] \qquad (8)$$

for any $K$-conserving changes $\Delta_K\rho(x)$ of $\rho(x)$ of $\int f(\rho(x))dx = K$ (i.e. for $\Delta\rho(x)$'s satisfying $\int f(\rho(x) + \Delta\rho(x))dx = K$), with

$$\lim_{\Delta_K\rho \to 0} \frac{o_K[\rho;\Delta_K\rho]}{\|\Delta_K\rho\|} = 0 \ . \qquad (8a)$$

Maintaining the linearity requirement for $F|_K[\rho;.]$, except for linear $K[\rho]$ constraints,

$$\int h(x)\rho(x)dx = L \ , \qquad (9)$$

is not directly possible, because $(\Delta_K\rho(x))_1 + (\Delta_K\rho(x))_2$ is not $K$-conserving in general; however, a corresponding requirement is ensured by writing $F|_K[\rho;.]$ in a form with 'built-in' linearity, as the form taken below. Following a similar way of defining a restricted Gâteaux derivative $G|_K[\rho;.]$ meets a serious problem, since in general, $\rho(x) + \varepsilon\Delta_K\rho(x)$ runs out of the set of $\rho(x)$'s of the given $K$ (for which problem Sec.V gives a resolution); however, for linear constraints Eq.(9), the restriction of the domain of $G[\rho;.]$ is applicable to get a $G|_L[\rho;.]$, since

$$\int h(x)\Delta_L\rho(x)dx = 0 \ . \qquad (10)$$

It is important to recognize that, contrary to the definition of $F$, the definition of $F|_K$ is not unique: if $F|_K[\rho;.]$ is a $K$-restricted Fréchet derivative at $\rho(x)$ then $F|_K[\rho;.] + \mu\int dx\, f^{(1)}(\rho(x)).$ , with any $\mu$, is that either, since

$$\lim_{\Delta_K \rho \to 0} \frac{\int f^{(1)}(\rho(x)) \Delta_K \rho(x)\,dx}{\|\Delta_K \rho\|} = 0 \qquad (11)$$

(or in the usual notation, $\int f^{(1)}(\rho(x)) \delta_K \rho(x)\,dx = 0$), following from $\int f^{(1)}(\rho) \Delta_K \rho = \int f(\rho + \Delta_K \rho) - \int f(\rho) + o[\rho; \Delta_K \rho] = o[\rho; \Delta_K \rho]$. Note that, because of Eq.(10), the ambiguity of $F|_K$ disappears for linear constraints, that is, $F|_L$ is unique.

In a physical relation, a functional derivative usually appears through

$$\frac{\delta A[\rho]}{\delta \rho(x)} := D[\rho(x'); \delta(x - x')] , \qquad (12)$$

with $\delta(x - x')$ the Dirac delta function, that is, $D[\rho; \cdot]$ is written as

$$D[\rho; \Delta \rho] = \int \frac{\delta A[\rho]}{\delta \rho(x)} \Delta \rho(x)\,dx , \qquad (13)$$

where $D[\rho; \cdot]$ can be the Fréchet or the Gâteaux derivative either. Keeping that representation (embodying the linearity of $D[\rho; \cdot]$) for restricted derivatives as well, a restricted Fréchet derivative or, for $K=L$, a restricted Gâteaux derivative is given as

$$D|_K[\rho; \Delta_K \rho] = \int \left.\frac{\delta A[\rho]}{\delta \rho(x)}\right|_K \Delta_K \rho(x)\,dx \qquad (14)$$

(for all $\Delta_K \rho(x)$'s). Now, the ambiguity of $F|_K[\rho; \cdot]$ appears as the ambiguity of $\left.\frac{\delta A[\rho]}{\delta \rho(x)}\right|_K$:

$$\left.\frac{\delta A[\rho]}{\delta \rho(x)}\right|_K + \mu f^{(1)}(\rho(x)) \qquad (15)$$

(with $\mu$ arbitrary), which, however, includes linear $f(\rho)$'s as well and holds even for Gâteaux $\left.\frac{\delta A[\rho]}{\delta \rho(x)}\right|_L$'s, because of Eq.(10).

## III. The most general form for constrained derivatives

With the definition of restricted derivatives, the *K*-conserving differentiation formula can be written as

$$\frac{\delta A[\rho]}{\delta_K \rho(x)} = \left.\frac{\delta A[\rho]}{\delta \rho(x)}\right|_K - \frac{f^{(1)}(\rho(x))}{K} \int \frac{f(\rho(x'))}{f^{(1)}(\rho(x'))} \left.\frac{\delta A[\rho]}{\delta \rho(x')}\right|_K dx' , \qquad (16)$$

since the chain rule

$$\frac{\delta A[\rho[g]]}{\delta g(x)} = \int \left.\frac{\delta A[\rho]}{\delta \rho(x')}\right|_K \frac{\delta \rho(x')}{\delta g(x)} dx' \qquad (17)$$

can be proved for functionals $\rho(x)[g]$ for which

$$\int f(\rho(x)[g]) dx = K \qquad (17a)$$

for any $g(x')$ [see Appendix], so the derivation of Eq.(1) holds also with $\left.\frac{\delta A[\rho]}{\delta \rho(x)}\right|_K$ (which of course can be $\frac{\delta A[\rho]}{\delta \rho(x)}$ if it exists) in the place of $\frac{\delta A[\rho]}{\delta \rho(x)}$. It can be seen that Eq.(16) yields a unique $\frac{\delta A[\rho]}{\delta_K \rho(x)}$, cancelling the ambiguity Eq.(15) of restricted derivatives (appearing in the form of "differentiational constants" $\mu$), since

$$\mu f^{(1)}(\rho(x)) - \frac{f^{(1)}(\rho(x))}{K} \int \frac{f(\rho(x'))}{f^{(1)}(\rho(x'))} \mu f^{(1)}(\rho(x')) dx' = 0 . \qquad (18)$$

Here and in the next Section, the question will be examined as in what sense the formula Eq.(16) can be considered as the one for *K*-conserving constrained differentiation.

To start with, consider the essential property [1,2] of $\dfrac{\delta}{\delta_K \rho}$ derivatives that $\dfrac{\delta A[\rho]}{\delta_K \rho(x)}$ gives $D[\rho; \Delta^*_K \rho]$ for the component

$$\Delta^*_K \rho(x) = \int \left\{ \delta(x - x') - \frac{1}{K} \frac{f(\rho(x))}{f^{(1)}(\rho(x))} f^{(1)}(\rho(x')) \right\} \Delta \rho(x') dx' \qquad (19)$$

of any $\Delta \rho(x)$ via

$$D[\rho; \Delta^*_K \rho] = \int \frac{\delta A[\rho]}{\delta_K \rho(x)} \Delta \rho(x) dx \ . \qquad (20)$$

Note that $\Delta^*_K \rho(x)$, for which

$$\int f^{(1)}(\rho(x)) \Delta^*_K \rho(x) dx = 0 \ , \qquad (21)$$

is not a *K*-conserving change in general; only first-order *K*-conserving variations $\delta_K \rho(x)$ satisfy (by definition) Eq.(21). The general (linear) form of a projection $\Delta \rho \to \Delta^*_K \rho$, with the requirement of being an identity for $\Delta^*_K \rho(x)$'s, is

$$\Delta^*_K \rho(x) = \int \left\{ \delta(x - x') - \frac{u(x)}{f^{(1)}(\rho(x))} f^{(1)}(\rho(x')) \right\} \Delta \rho(x') dx' \ , \qquad (22)$$

with

$$\int u(x) dx = 1 \ , \qquad (23a)$$

that is,

$$u(x) = \frac{q(x)}{\int q(x') dx'} \qquad (23b)$$

(for $\int q(x) dx \neq 0$), which yields

$$\int \left. \frac{\delta A[\rho]}{\delta \rho(x)} \right|_K \Delta^*_K \rho(x) dx = \int \left\{ \left. \frac{\delta A[\rho]}{\delta \rho(x)} \right|_K - f^{(1)}(\rho(x)) \int \frac{u(x')}{f^{(1)}(\rho(x'))} \left. \frac{\delta A[\rho]}{\delta \rho(x')} \right|_K dx' \right\} \Delta \rho(x) dx \qquad (24)$$

(which gives $D|_K[\rho;\delta_K\rho]$ for $\Delta^*_K\rho(x) = \delta_K\rho(x)$).

The formula

$$\frac{\delta A[\rho]}{\delta'_K\rho(x)} = \left.\frac{\delta A[\rho]}{\delta\rho(x)}\right|_K - f^{(1)}(\rho(x))\int \frac{u(x')}{f^{(1)}(\rho(x'))}\left.\frac{\delta A[\rho]}{\delta\rho(x')}\right|_K dx' \qquad (25)$$

fulfils the most essential requirement on a proper treatment of constrained functional differentiation; namely, for functionals that are equal on a set of $\rho(x)$'s restricted by Eq.(2), whose restricted derivatives $\left.\frac{\delta}{\delta\rho(x)}\right|_K$ therefore may differ only by a $cf^{(1)}(\rho(x))$, i.e.

$$\left.\frac{\delta A[\rho_K]}{\delta\rho(x)}\right|_K - \left.\frac{\delta B[\rho_K]}{\delta\rho(x)}\right|_K = cf^{(1)}(\rho(x)) , \qquad (26)$$

formula (25) gives equal derivatives,

$$\frac{\delta A[\rho_K]}{\delta'_K\rho(x)} = \frac{\delta B[\rho_K]}{\delta'_K\rho(x)} , \qquad (27)$$

cancelling any $cf^{(1)}(\rho(x))$. From that property then the naturally expectable

$$\frac{\delta b(K[\rho])}{\delta'_K\rho(x)} = 0 \qquad (28)$$

follows straight, with arbitrary function $b(K)$, since $\frac{\delta b(K[\rho])}{\delta\rho(x)} = \frac{\partial b(K)}{\partial K}f^{(1)}(\rho(x))$.

Requiring the property Eq.(27), in itself leads to the formula Eq.(25) as well, since Eq.(26) gives

$$\int \frac{u(x)}{f^{(1)}(\rho(x))}\left.\frac{\delta A[\rho]}{\delta\rho(x)}\right|_K dx - \int \frac{u(x)}{f^{(1)}(\rho(x))}\left.\frac{\delta B[\rho]}{\delta\rho(x)}\right|_K dx = c$$

with an $u(x)$ satisfying Eq.(23) (as the operator to act on both sides of Eq.(26) has to be linear to get a proper $c$, dividable into two parts for the two $\dfrac{\delta}{\delta'_K \rho}$ derivatives), yielding Eq.(25).

$u(x)$ in Eq.(25) can be both some functional of $\rho(x)$ and a $\rho(x)$-independent function. A possible choice, e.g., is $u(x) = \delta(x - x_0)$, leading to

$$\frac{\delta A[\rho]}{\delta'_K \rho(x)} = \left.\frac{\delta A[\rho]}{\delta \rho(x)}\right|_K - f^{(1)}(\rho(x)) \frac{1}{f^{(1)}(\rho(x_0))} \left.\frac{\delta A[\rho]}{\delta \rho(x_0)}\right|_K . \qquad (29)$$

That the main essence of constrained differentiation is embodied by the general form Eq.(25), gives some more "understanding" of why, in the formula Eq.(16), the multiplyer of $\left.\dfrac{\delta A[\rho]}{\delta \rho(x)}\right|_K$ in the integrand remains undifferentiated in getting constrained second derivatives, as shown through the example of number-conserving derivatives (where $f(\rho) = \rho$) in [4].

### IV. Requirement from independence of constraint

So far, in Secs.II-III, the concrete form Eq.(2) of constraints has not been utilized, hence the general $\dfrac{\delta C[\rho]}{\delta \rho(x)}$, emerging from the general constraint form $C[\rho] = 0$, can be written in the place of $f^{(1)}(\rho(x))$ in all expressions there. For constraints Eq.(2), however, a mathematically meaningful choice of $u(x)$ in Eq.(25) arises, giving the formula Eq.(16).

In the case of number ($N$) conservation, $f(\rho) = \rho$ [see remark concerning the origin of the name later], the choice $q(x) = \rho(x)$ yields a fixation of $\left.\dfrac{\delta A[\rho]}{\delta \rho(x)}\right|_N$ that is determined by its weighted average $\displaystyle\int \dfrac{\rho(x)}{\int \rho(x')dx'} \left.\dfrac{\delta A[\rho]}{\delta \rho(x)}\right|_N dx$. Why that choice is special mathematically is that for functionals $A[\rho]$ homogeneous of degree zero, that is,

$$A[\lambda\rho] = A[\rho] \tag{30}$$

for any $x$-independent $\lambda$, it (that is, Eq.(16), with $f(\rho) = \rho$) gives

$$\frac{\delta A[\rho]}{\delta_N \rho(x)} = \frac{\delta A[\rho]}{\delta \rho(x)}, \tag{31}$$

since from the homogeneity Eq.(30),

$$\int \rho(x) \frac{\delta A[\rho]}{\delta \rho(x)} dx = 0 \tag{32}$$

follows (if $A[\rho]$ is differentiable). A degree-zero homogeneous functional, on the other hand, is special since it is independent of $N$ (and vice versa), as $A[\rho] = A[\dfrac{\rho}{\int \rho}\int \rho] = A[\dfrac{\rho}{\int \rho}]$, and for $N$-independent functionals, an $N$-conservation constraint yields no restriction on their variations, therefore Eq.(31) is expectable, that is the $N$-constraint has no effect on their differentiation. (Note that Eq.(31) for degree-zero homogeneous $A[\rho]$, as a condition, along with the general form Eq.(25), gives back $q(x) = \rho(x)$, that is, that choice is equivalent with the requirement of Eq.(31).)

Thus, for $f(\rho) = \rho$, the formula Eq.(16) is superior mathematically over other forms Eq.(25); but for other constraints Eq.(2), can Eq.(16) still be considered as special among the forms Eq.(25)? Can an independence of $K$, in general, be understood? Contrary to the case of the N-constraint, where the N-dependence could be naturally separated as $\rho(x) = \left(\int \rho(x')\right) \frac{\rho(x)}{\int \rho(x')}$, there is no trivial intuitive way to answer this question of separation of K in $\rho(x)$. The following generalization of the concept of homogeneity of degree m, however, gives a natural solution:

$$A[f^{-1}(\lambda f(\rho))] = \lambda^m A[\rho] \ . \tag{33}$$

Eq.(33) can be considered as a generalized, deformed homogeneity (K-homogeneity) of degree m, and yields

$$\int \frac{f(\rho(x))}{f^{(1)}(\rho(x))} \frac{\delta A[\rho]}{\delta \rho(x)} dx = m A[\rho] \ . \tag{34}$$

In this way, an understanding of the particular choice $q(x) = f(\rho(x))$ is obtained, the formula Eq.(16) giving

$$\frac{\delta A[\rho]}{\delta_K \rho(x)} = \frac{\delta A[\rho]}{\delta \rho(x)} \tag{35}$$

for K-independent (that is, by definition, degree-zero K-homogeneous) functionals $A[\rho]$, where the K-constraint is expectable to have no effect on differentiation.

The requirements Eq.(27) and Eq.(35) practically mean that the K-conserving derivative of a functional $A[\rho]$ at a $\rho_K(x)$, as given by Eq.(16), is defined as the unrestricted derivative of the degree-zero homogeneous extension of $A[\rho]$,

$$A_K^0[\rho] := A\left[f^{-1}\left(\frac{K}{\int f(\rho(x))\, dx} f(\rho)\right)\right], \tag{36}$$

at $\rho_K(x)$, on the basis of

$$\frac{\delta A[\rho_K]}{\delta_K \rho(x)} = \frac{\delta A_K^0[\rho_K]}{\delta_K \rho(x)} = \frac{\delta A_K^0[\rho_K]}{\delta \rho(x)} \ , \tag{37}$$

where only Eqs.(27) and (35) have been applied, utilizing the two special properties of $A_K^0[\rho]$, namely, that it gives $A[\rho]$ for $\rho_K(x)$'s (that is, $\rho(x)$'s of $\int f(\rho(x))dx = K$), and it is independent of $K$. Thus,

$$\frac{\delta A[\rho]}{\delta_K \rho(x)} := \frac{\delta A[f^{-1}(\frac{K}{\int f(\rho)} f(\rho))]}{\delta \rho(x)} , \qquad (38)$$

which yields the formula Eq.(16) (via the chain rule Eq.(17) applied for Eq.(36)), and sums up the mathematical basis behind the derivation of Eq.(1) in [1,2] (see the application of Eq.(4), with the decomposition Eq.(3)). In connection with this approach Eq.(38), it is important to note that the extension

$$\rho_K^0[\rho] = f^{-1}\left(\frac{K}{\int f(\rho(x'))dx'} f(\rho(x))\right) \qquad (39)$$

from a $\rho_K(x)$ is unique, that is, it is the only extension that (i) reduces to $\rho_K(x)$ for $\rho_K(x)$ and (ii) is independent of $K$, which can also be seen by the following: Any $\rho(x)$ can be written as $\rho(x) = f^{-1}(\frac{\int f(\rho)}{K} f(\rho_K(x)))$ [with some uniquely determined $\rho_K(x)$: $\rho_K(x) = f^{-1}(\frac{K}{\int f(\rho)} f(\rho(x)))$], therefore for an $A_K^{0*}[\rho]$ satisfying the two conditions $A_K^{0*}[\rho_K] = A[\rho_K]$ (for any $\rho_K(x)$) and independence of $K$,

$$A_K^{0*}[\rho] = A_K^{0*}[f^{-1}(\frac{\int f(\rho)}{K} f(\rho_K))] = A_K^{0*}[\rho_K] = A[\rho_K] = A[f^{-1}(\frac{K}{\int f(\rho)} f(\rho))] , \qquad (40)$$

that is $A_K^{0*}[\rho]$ is unique (for any $A[\rho]$, including $A[\rho] = \rho$), and is $A_K^0[\rho]$. For that unique extension $\rho_K^0[\rho]$,

$$\int f(\rho_K^0[\rho](x))dx = K \qquad (41)$$

also holds (for any $\rho(x)$), which was used as a basic requirement in [2] to obtain the proper decomposition Eq.(3) (which, with $g(x) = \rho(x)$, gives $\rho_K^0[\rho]$).

## V. *K*-conserving derivative as the complementer of a derivative with respect to *K*

For $f(\rho) = \rho$, the superiority of Eq.(16) (yielded by the requirement given by Eqs.(30) and (31)) among the forms Eq.(25), is also shown by the fact that an *N*-conserving derivative emerges as

$$\frac{\delta A[\rho]}{\delta_N \rho(x)} = \frac{\delta A[\rho]}{\delta \rho(x)} - \left(\frac{\partial A[Nn]}{\partial N}\right)_{n(x')} \tag{42}$$

(with the shape $n(x)$ of $\rho(x)$ defined as $n(x) := \dfrac{\rho(x)}{\int \rho(x')\,dx'}$, that is, $\rho(x) = Nn(x)$), since

$$\left(\frac{\partial A[N n(x')]}{\partial N}\right)_{n(x')} = \int \frac{\delta A[\rho(x')]}{\delta \rho(x)}\left(\frac{\partial \rho(x)}{\partial N}\right)_{n(x)} dx = \int \frac{\delta A[\rho(x')]}{\delta \rho(x)}\frac{\rho(x)}{N}\,dx \ . \tag{43}$$

Hence,

$$\frac{\delta A[\rho]}{\delta \rho(x)} = \frac{\delta A[\rho]}{\delta_N \rho(x)} + \left(\frac{\partial A[Nn]}{\partial N}\right)_{n(x')}, \tag{44}$$

by which a differential $D[\rho;\Delta\rho] = \int \dfrac{\delta A[\rho]}{\delta \rho(x)}\Delta\rho(x)\,dx$ splits as

$$\int \frac{\delta A[\rho]}{\delta \rho(x)}\Delta\rho(x)\,dx = \int \frac{\delta A[\rho]}{\delta_N \rho(x)}\Delta_N \rho(x)\,dx + \left(\frac{\partial A[Nn]}{\partial N}\right)_{n(x')}\Delta N \ , \tag{45}$$

two components vanishing in Eq.(45). The derivative Eq.(43) can be identified as the shape-conserving derivative [1]

$$\frac{\delta A[\rho]}{\delta_{n(x')}\rho(x)} \equiv \frac{1}{N}\int \rho(x)\frac{\delta A[\rho]}{\delta\rho(x)}dx \ . \tag{46}$$

Eq.(44) can alternatively be written as

$$\frac{\delta A[\rho]}{\delta\rho(x)} = \frac{1}{N}\left(\frac{\delta A[Nn]}{\delta_1 n(x)}\right)_N + \left(\frac{\partial A[Nn]}{\partial N}\right)_{n(x')} , \tag{47}$$

using

$$\frac{1}{N}\left(\frac{\delta A[Nn]}{\delta_1 n(x)}\right)_N = \frac{1}{N}\int \frac{\delta A[\rho]}{\delta\rho(x')}N\frac{\delta n(x')}{\delta_1 n(x)}dx' = \frac{\delta A[\rho]}{\delta_N \rho(x)} , \tag{48}$$

or directly through

$$\frac{\delta A[Nn]}{\delta\rho(x)} = \int\left(\frac{\delta A[Nn]}{\delta n(x')}\right)_N \frac{\delta n(x')}{\delta\rho(x)}dx' + \left(\frac{\partial A[Nn]}{\partial N}\right)_n \frac{\delta N}{\delta\rho(x)} \ .$$

Eq.(47) represents a conceptual alternative to

$$\frac{\delta A[\rho]}{\delta\rho(x)} = \frac{\delta A[\rho]}{\delta_N \rho(x)} + \frac{\delta A[\rho]}{\delta_{n(x')}\rho(x)} \ . \tag{49}$$

The above decomposition (Eqs.(44), (47) and (49)) of a functional derivative can be generalized for linear $K$-constraints, Eq.(9), through decomposing the functional variable as $\rho(x) = Ll(x)$, with the "$L$-shape" $l(x)$ of $\rho(x)$ defined as

$$l(x) := \frac{\rho(x)}{\int h(x')\rho(x')dx'} \tag{50}$$

(that is, $\int h(x)l(x)dx = 1$), yielding

$$\frac{\delta A[\rho]}{\delta\rho(x)} = \frac{\delta A[\rho]}{\delta_L \rho(x)} + \left(\frac{\partial A[Ll]}{\partial L}\right)_{l(x')} , \tag{51}$$

and the other corresponding relationships. The further generalization embracing an arbitrary $K$-constraint can only be done formally, there being no meaning of "differentiating with respect to $K$ of the functional variable *only*, while conserving the other part of the variable" (except for homogeneous $K[\rho]$'s, leading to the concept of "H-shape").

The decomposition of a functional derivative into a number-conserving and a shape-conserving part gets special conceptual relevance with respect to the density-shape based reformulation [5] of the density functional theory (DFT) of many-electron systems [6] into a "density-shape functional theory". (The name "number-conserving derivative" was taken from an "unofficial" use by some theoreticians in DFT, where the fixation of the particle number $N$ is needed, to term $N$-restricted derivatives.) With the use of $N$-conserving differentiation, the following relationship connecting the derivative of a density-shape functional $A'[n] \equiv A[\rho[n]]$ and the derivative of the corresponding density functional $A[\rho]$ arises:

$$\frac{\delta A'[n]}{\delta_1 n(x)} = \frac{\delta A[N[n]n]}{\delta_1 n(x)} = \left(\frac{\partial A[Nn]}{\partial N}\right)_{n(x')} \frac{\delta N}{\delta_1 n(x)} + \frac{\delta A[\rho]}{\delta_N \rho(x)} N =$$

$$= N \frac{\delta A[\rho]}{\delta \rho(x)}\bigg|_{(+/-)} + \left(\frac{\partial A[Nn]}{\partial N}\bigg|_{(+/-)}\right)_{n(x')} \left(\frac{\delta N}{\delta_1 n(x)} - N\right), \quad (52a)$$

or

$$\frac{\delta A'[n]}{\delta_1 n(x)} = \frac{\delta A[\rho]}{\delta_{n(x')} \rho(x)} \frac{\delta N}{\delta_1 n(x)} + \frac{\delta A[\rho]}{\delta_N \rho(x)} N =$$

$$= N \frac{\delta A[\rho]}{\delta \rho(x)}\bigg|_{(+/-)} + \frac{\delta A[\rho]}{\delta_{n(x')} \rho(x)}\bigg|_{(+/-)} \left(\frac{\delta N}{\delta_1 n(x)} - N\right). \quad (52b)$$

(The last equality in Eq.(52) is for density functionals $A[\rho]$ having an unconstrained derivative, or at least, right/left-side derivative $\frac{\delta A[\rho]}{\delta \rho(x)}\bigg|_{+/-}$ .) Note that the account for $N$- conservation is essential in the case of density-shape-functional derivatives, as no meaning is associated to a change in norm of $n(x)$.

The derivative $\left(\dfrac{\partial E_v[Nn]}{\partial N}\right)_{n(x)}$ of the ground-state energy density functional, for the ground-state density corresponding to a given external potential $v(x)$, is closely related to the derivative of the ground-state energy $E^{gs}(N,v)$ with respect to the particle number $N$, namely, they are equal, since

$$\left(\frac{\partial E_v[Nn]}{\partial N}\right)_{n(x')} = \int \frac{\delta E_v[\rho]}{\delta \rho(x)} \frac{\rho(x)}{N} dx = \int \mu \frac{\rho(x)}{N} dx = \mu = \frac{\partial E^{gs}(N,v)}{\partial N} \;, \qquad (53)$$

using the Euler equation for the ground-state density of density-functional theory in the second equality.

Finally in this Section, the opportunity is taken to note an incorrectness in [1]: It is stated in [1] (in the first sentence of the last paragraph on page 2) that there is a false view in density-functional theory that the *N*-conserving derivative of a functional is determined only up to an arbitrary constant (in *x*), while that ambiguity is, correctly, referred (in DFT) to the *N*-restricted derivative. [That was an unfortunate reaction by the author to the opinion of some DFT theoreticians (not necessarily those referenced in the last paragraph on page 2) about the *N*-conserving derivative formula that a unique derivative over an *N*-restricted domain is a nonsense.]

**VI. Constrained derivatives as deformed Gâteaux derivatives**

As pointed out in Sec.II, the Gâteaux definition of unrestricted derivatives cannot simply be restricted to obtain a definition for a *K*-restricted derivative. First, it would be a senseless definition, since the straight path $\rho_K(x) + \varepsilon \Delta_K \rho(x)$ runs out of the restricted domain of $\rho_K(x)$'s (except for $K=L$), which gets particular relevance in the case of functionals defined only over the set of $\rho_K(x)$'s. Second, the essential connection Eq.(7) between Gâteaux and Fréchet derivatives would not apply for restricted derivatives (as $A[\rho + \varepsilon \Delta_K \rho] - A[\rho]$ cannot be given as $F|_K[\rho; \varepsilon \Delta_K \rho] - o_K[\rho; \varepsilon \Delta_K \rho]$), that is, the existence of K-restricted Fréchet derivative would not mean anything for the corresponding restricted Gâteaux derivative.

To treat the above problem, that is, to have a proper restricted Gâteaux derivative concept, a generalization of the concept of the Gâteaux derivative is necessary, getting to a generalized, deformed Gâteaux derivative. For that, the following deformation of the linear path $\rho(x) + \varepsilon \Delta_K \rho(x)$ needs to be introduced:

$$f^{-1}\left(\frac{K}{\int f(\rho(x') + \varepsilon \Delta_K \rho(x'))\, dx'} f(\rho(x) + \varepsilon \Delta_K \rho(x))\right), \qquad (54)$$

which yields a $\rho_K(x)$ for any value of $\varepsilon$, and for a $\rho_K(x)$ (that is, for $\rho(x)$ at $\varepsilon = 0$, and for $\rho(x) + \Delta_K \rho(x)$ at $\varepsilon = 1$), reduces to it. With the above general K-conserving path, then the following K-restriction of the Gâteaux derivative emerges:

$$G|_K[\rho; \Delta_K \rho] := \lim_{\varepsilon \to 0} \frac{A[f^{-1}(\frac{K}{\int f(\rho + \varepsilon \Delta_K \rho)} f(\rho + \varepsilon \Delta_K \rho))] - A[\rho]}{\varepsilon}, \qquad (55)$$

which is coherent conceptually, and has the proper connection to the K-restricted Fréchet derivative, that is, exists if $F|_K[\rho; .]$ exists:

$$G|_K[\rho; \Delta_K \rho] = \lim_{\varepsilon \to 0} \frac{1}{\varepsilon}\left(\int \left.\frac{\delta A[\rho]}{\delta \rho(x)}\right|_K \{\rho_K^0[\rho + \varepsilon \Delta_K \rho](x) - \rho(x)\}\, dx + o_K[\rho; \{\ldots\}]\right) =$$

$$\lim_{\varepsilon \to 0} \frac{1}{\varepsilon} \left( \int \left. \frac{\delta A[\rho]}{\delta \rho(x)} \right|_K \left\{ \left. \frac{d \rho_K^0[\rho + \varepsilon \Delta_K \rho](x)}{d \varepsilon} \right|_{\varepsilon=0} \varepsilon + o(0;\varepsilon) \right\} dx + o_K[\rho;\{\ldots\}] \right) =$$

$$\int \left. \frac{\delta A[\rho]}{\delta \rho(x)} \right|_K \left\{ \int \frac{\delta \rho_K^0[\rho](x)}{\delta \rho(x')} \Delta_K \rho(x') \, dx' + \lim_{\varepsilon \to 0} \frac{o(0;\varepsilon)}{\varepsilon} \right\} dx +$$

$$\lim_{\varepsilon \to 0} \left( \frac{\| \rho_K^0[\rho + \varepsilon \Delta_K \rho] - \rho \|}{\varepsilon} \frac{o_K[\rho;\{\rho_K^0[\rho + \varepsilon \Delta_K \rho] - \rho\}]}{\| \rho_K^0[\rho + \varepsilon \Delta_K \rho] - \rho \|} \right) =$$

$$\iint \left. \frac{\delta A[\rho]}{\delta \rho(x)} \right|_K \frac{\delta \rho(x)}{\delta_K \rho(x')} dx \, \Delta_K \rho(x') \, dx' \tag{56}$$

($\| \rho_K^0[\rho + \varepsilon \Delta_K \rho] - \rho \| / \varepsilon$ being bounded in $\varepsilon$), giving

$$\left. G \right|_K [\rho; \Delta_K \rho] = \int \frac{\delta A[\rho]}{\delta_K \rho(x)} \Delta_K \rho(x) \, dx \ . \tag{57}$$

Eq.(57) constitutes a strong mathematical basis for the *K*-conserving derivative Eq.(16) (and gives an intuitive interpretation), showing that it emerges as a Gâteaux (that is, directional) derivative over the restricted set of $\rho_K(x)$'s. Also, the above origination of $\frac{\delta}{\delta_K \rho}$ throws more light on the mathematical basis of the derivation of the formula Eq.(16) in [1,2], that is, on the decomposition Eq.(3). Note that Eq.(57) determines $\frac{\delta A[\rho]}{\delta_K \rho(x)}$ uniquely (contrary to the definition of a restricted Fréchet derivative), except for linear constraints, for which Eq.(10). It follows from the above theorem as well that $\left. G \right|_K$ exists more generally than $\left. F \right|_K$, similarly to the case of the unrestricted Gâteaux and Fréchet derivative, indicating the relevance of Eq.(55) for physical applications, the Fréchet definition for a derivative being too strong for a general applicability in physics.

It is important to point out that the generalization of the Gâteaux derivative for nonlinear (*K*-) paths is possible only on the *K*-restricted sets (and not on the whole set of $\rho(x)$'s), contrary to its special case for *K=L*, that is, the (original) Gâteaux derivative $G[\rho;.]$ (which is defined on the set of all $\Delta\rho(x)$'s, not only for $\Delta_L\rho(x)$'s). For, the existence of a *K*-deformed Gâteaux derivative on an unrestricted set leads to a contradiction, as on one hand, an unrestricted *K*-deformed Gâteaux derivative should have the form as in Eq.(57) because of linearity in $\Delta\rho(x)$, and on the other hand, for linear constraints (where it should reduce to $G[\rho;.]$), should equal the (unrestricted) Fréchet derivative, if that exists. [Only for linear constraints, the difference between $\frac{\delta A[\rho]}{\delta\rho(x)}$ and $\frac{\delta A[\rho]}{\delta_K\rho(x)}$ cancels out in Eq.(57) (because of Eq.(10)), 'allowing' the existence of an unrestricted *L*-deformed Gâteaux derivative, which is just $G[\rho;.]$]. That nonexistence of an unrestricted *K*-deformed Gâteaux derivative can be interpreted as the impossibility of rotating a *K*-conserving path around a $\rho_K(x)$ without deformation to get a corresponding path for a general $\Delta\rho(x)$, except for *L*-conserving (i.e. straight) paths.

It has to be emphasized here that the ambiguity of restricted Fréchet derivatives can also be fixed as Eq.(16), obtaining the concept of a constrained Fréchet derivative (and completing the analogy with Eq.(7), $F|_K[\rho;.]$ giving directly $G|_K[\rho;.]$ in this way).

Finally, it is important to underline that the two originations of the *K*-conserving derivative Eq.(16) given in the present and the preceding Sections apply for invertible $f(\rho)$'s, noting though that independence of $K$ could be defined by the more general Eq.(34), with $m=0$, as well.

## VII. Summary

The mathematical basics of $K$-conserving functional differentiation, and of the $K$-conserving derivative formula Eq.(16), have been clarified, which is essential for its physical use (as [7], in the modelling of simultaneous dewetting and phase separation in thin liquid films [8]), for its conceptual application (see [1,4] in density-functional theory [6], and Sec.V), and for its generalization for wider classes of constraints (as [9] for simultaneous $K$-constraints).

On the basis of the most substantial requirement a (properly defined) $K$-conserving derivative has to satisfy, the general form Eq.(25) for them has been derived. Showing the superiority of the formula Eq.(16) over other forms Eq.(25) in the case of $N$-conservation (the simplest form of the $K$-constraint Eq.(2)), it has been pointed out that for $N$-independent functionals, Eq.(16) yields the unconstrained derivative, in accordance with natural expectations. Via a generalization of the concept of homogeneity of functionals, then the same arises for $K$-independent functionals, that is, the $K$-constraint has no effect on the differentiation of them. These results yield a method for generalizations of $K$-conserving differentiation [9].

A $K$-conserving derivative can be considered as a part of the unconstrained, that is, full, derivative, the other part of which has been shown (in Sec.V) to be a simple derivative with respect to $K$ for linear $K$-constraints. Connections with the derivative with respect to the "shape" of the functional variable and with the shape-conserving derivative, together with their use in density-functional theory [6], are also discussed in Sec.V.

Yielding an intuitive interpretation of *K*-conserving derivatives, it has also been shown that *K*-conserving derivatives can be originated as directional derivatives, via a generalization of the Gâteaux derivative for nonlinear *K*-conserving paths, Eq.(55).

Finally, a remark on terminology: A derivative has been termed "constrained" if it does not emerge through a simple restriction of domain of validity but has a modified definition, to handle a constraint properly; hence it cannot simply be chosen to be the unrestricted derivative if that exists (contrary to a restricted Fréchet derivative, e.g.).

## Appendix: Proof of Eq.(17)

The chain rule Eq.(17) can be proved in the following way for Fréchet derivatives:

$$A[\rho[g+\Delta g]] - A[\rho[g]] = \int \left.\frac{\delta A}{\delta \rho(x)}\right|_K [\rho[g]]\{\rho[g+\Delta g](x) - \rho[g](x)\}dx + o_K^A[\rho[g];\{...\}] =$$

$$\int \left.\frac{\delta A}{\delta \rho(x)}\right|_K [\rho[g]]\left\{\int \frac{\delta \rho[g](x)}{\delta g(x')}\Delta g(x')dx' + o^{\rho(x)}[g;\Delta g]\right\}dx + o_K^A[\rho[g];\{...\}] =$$

$$\iint \left.\frac{\delta A[\rho]}{\delta \rho(x)}\right|_K \frac{\delta \rho[g](x)}{\delta g(x')}dx\,\Delta g(x')dx' + \int \left.\frac{\delta A[\rho]}{\delta \rho(x)}\right|_K o^{\rho(x)}[g;\Delta g]\,dx + o_K^A[\rho[g];\{...\}] \,, \quad (A1)$$

with

$$\lim_{\Delta g \to 0} \frac{o_K^A[\rho[g];\{\rho[g+\Delta g]-\rho[g]\}]}{\|\rho[g+\Delta g]-\rho[g]\|} \frac{\|\rho[g+\Delta g]-\rho[g]\|}{\|\Delta g\|} = 0$$

and

$$\lim_{\Delta g \to 0} \frac{o^{\rho(x)}[g;\Delta g]}{\|\Delta g\|} = 0 \; ,$$

noting that $\rho[g+\Delta g]-\rho[g]$ is a $K$-conserving change in $\rho(x)$, according to Eq.(17a).

It is important that $\int \left.\frac{\delta A[\rho]}{\delta \rho(x')}\right|_K \frac{\delta \rho(x')}{\delta g(x)} dx'$ is unique since the ambiguity $+\mu f^{(1)}(\rho(x))$ of $\left.\frac{\delta A[\rho]}{\delta \rho(x)}\right|_K$ is cancelled in that expression, as

$$\int \mu f^{(1)}(\rho[g](x')) \frac{\delta \rho[g](x')}{\delta g(x)} dx' = \mu \frac{\delta \int f(\rho[g](x'))dx'}{\delta g(x)} = \mu \frac{\delta K}{\delta g(x)} = 0 \quad (A2)$$

because of (17a).

Eq.(17) also holds if $\frac{\delta \rho[g](x')}{\delta g(x)}$ is only a Gâteaux derivative; then $\frac{\delta A[\rho[g]]}{\delta g(x)}$ also is a Gâteaux derivative. For,

$$\lim_{\varepsilon \to 0} \frac{A[\rho[g+\varepsilon \Delta g]] - A[\rho[g]]}{\varepsilon} =$$

$$\lim_{\varepsilon \to 0} \frac{1}{\varepsilon} \left( \int \left.\frac{\delta A}{\delta \rho(x)}\right|_K [\rho[g]] \{\rho[g+\varepsilon \Delta g](x) - \rho[g](x)\} dx + o_K^A[\rho[g];\{...\}] \right) =$$

$$\int \left.\frac{\delta A[\rho]}{\delta \rho(x)}\right|_K \int \frac{\delta \rho[g](x)}{\delta g(x')} \Delta g(x') dx' \, dx + \lim_{\varepsilon \to 0} \frac{\|\rho[g+\varepsilon \Delta g]-\rho[g]\|}{\varepsilon} \frac{o_K^A[\rho[g];\{\rho[g+\varepsilon \Delta g]-\rho[g]\}]}{\|\rho[g+\varepsilon \Delta g]-\rho[g]\|} =$$

$$\iint \left.\frac{\delta A[\rho]}{\delta \rho(x)}\right|_K \frac{\delta \rho[g](x)}{\delta g(x')} dx \, \Delta g(x') \, dx' \; , \tag{A3}$$

and Eq.(A2).

For *K=L*, Eq.(17) is valid for Gâteaux derivatives (all derivatives in Eq.(17) are Gâteaux, embracing the case, of course, in which $\frac{\delta\rho[g](x')}{\delta g(x)}$ is Fréchet) as well, in the proof of which also (A2) is the new element, compared to the unconstrained case.

**Acknowledgments:** Grant D048675 from OTKA is gratefully acknowledged.

E-mail address: galt@phys.unideb.hu

[1] T. Gál, Phys. Rev. A 63 (2001) 022506; *ibid.* 74 (2006) 029901 (Erratum).

[2] T. Gál, J. Phys. A 35 (2002) 5899.

[3] For the detailed definition of the Fréchet derivative and of the Gâteaux derivative, see e.g. P. Blanchard and E. Brüning, *Mathematical Methods in Physics* (Birkhäuser, Berlin, 2003).

[4] T. Gál, Phys. Lett. A 355 (2006) 148.

[5] R.G. Parr, J. Bartolotti, J. Phys. Chem. 87 (1983) 2810;

P.W. Ayers, Proc. Natl. Acad. Sci. USA 97 (2000) 1959;

P. Geerlings, F. De Proft, PW Ayers, In: *Theoretical and Computational Chemistry*, Vol. 19, ed. A. Toro-Labbé (Elsevier, Amsterdam, 2006), in press.

[6] P. Geerlings, F. De Proft, W. Langenaeker, Chem. Rev. 103 (2003) 1793, and references therein; see p.1814 on the shape (function) of density-functional theory.

[7] N. Clarke, Macromolecules 38 (2005) 6775.

[8] R. Yerushalmi-Rozen, T. Kerle, J. Klein, Science 285 (1999) 1254;

H. Wang, R. J. Composto, J. Chem. Phys. 113 (2000) 10386.

[9] T. Gál, arXiv:physics/0603129 (2006).